\documentclass{article}
\usepackage{spconf,amsmath,graphicx}
\usepackage{multirow}
\usepackage{enumitem}
\setlist{nosep, leftmargin=14pt}

\usepackage{mwe} % to get dummy images
\usepackage{algorithm}
\usepackage{algorithmic}
\title{INR-LDDMM: Fluid-based Medical Image Registration Integrating Implicit Neural Representation and Large Deformation Diffeomorphic Metric Mapping}
\name{Chulong Zhang, Xiaokun Liang}
\address{Shenzhen Institute of Advanced Technology, Chinese Academy of Sciences.}
\begin{document}
%\ninept
%
\maketitle
\begin{abstract}
We propose a fluid-based registration framework of medical images based on implicit neural  representation. By integrating implicit neural representation and Large  Deformable Diffeomorphic Metric Mapping (LDDMM), we employ a Multilayer  Perceptron (MLP) as a velocity generator while optimizing velocity and  image similarity. Moreover, we adopt a coarse-to-fine approach to  address the challenge of deformable-based registration methods dropping  into local optimal solutions, thus aiding the management of significant  deformations in medical image registration. Our algorithm has been  validated on a paired CT-CBCT dataset of 50 patients,taking the Dice coefficient of transferred annotations as an evaluation metric.  Compared to existing methods, our approach achieves the  state-of-the-art  performance.
\end{abstract}
\section{Introduction}
\label{sec:intro}

Medical  image  registration  is  a  critical  part  of  medical  image  processing,  aiding  in  the  collection  and  integration  of  information  from  various  image  sources.  The  registration  process  involves  modifying  the  geometric  configuration  of  one  image  to  align  with  another,  thereby  establishing  a  mapping  relationship  between  the  respective  points.  This  is  indispensable  for  many  medical  image  processing  applications,  such  as  in  Image  Guided  Radiation  Therapy  (IGRT).

The  techniques  for  volumetric  image  registration  can  be  mainly  categorized  into  two:  traditional  iterative-based  methods  and  deep  learning  methods.  Traditional  iterative-based  methods  such  as  bUnwarpJ \cite{b19}, NiftyReg \cite{b20}, RVSS \cite{b19}, ANTs \cite{b21}, DROP \cite{b22}, and Elastix \cite{b23},  hold  the  advantage  of  optimizing  each  image  individually;  however,  typically  they  are  slower,  susceptible  to  getting  stuck  in  local  optima,  with  complicated  parameter  optimization.  Traditional  iterative  methods  can  further  be  split  into  displacement  field  optimized  registration  methods  and  velocity  field  optimized  registration  methods. 

Displacement  field  optimized  methods,  often  referred  to  as  elastic  registrations,  include  techniques  like  Demos.  Velocity  field  optimized  methods,  commonly  known  as  Fluid  registration  methods,  often  possess  superior  diffeomorphism  and  physical  properties  (for  example,  better  intermediate  state  images  can  be  obtained).  Examples  of  Fluid  registration  methods  include  vSVF  and  LDDMM\cite{cao2005large}.  vSVF  makes  the  assumption  that  velocity  doesn’t  transform  with  time,  aiming  to  search  for  the  initial  momentum.  LDDMM  considers  the  speed  $v(x,t)$  as  a  function  of  position  and  time,  optimizing  for  image  similarity  and  speed,  with  the  intent  to  seek  the  best  initial  momentum.

Deep  learning-based  methods  rose  to  prominence  with  voxelmorph\cite{balakrishnan2019voxelmorph},  which  is  an  unsupervised  learning-based  approach  for  deformable  medical  image  registration.  It  uses  a  Convolutional  Neural  Network  (CNN)  to  approximate  the  registration  function  by  minimizing  image  dissimilarity  and  a  regularization  term  that  encourages  smooth  spatial  transformations.  However,  the  performance  of  deep  learning-based  methods  highly  depends  on  the  dataset,  and  performs  poorly  when  the  dataset  is  scarce  or  there  is  a  large  variance  between  training  and  test  datasets.

Implicit  Neural  Representation  is  an  emerging  technology\cite{ulyanov2018deep,mescheder2019occupancy,mildenhall2021nerf}.  It  uses  neural  networks  to  implicitly  represent  complex  functions,  shapes  and  structures  in  high-dimensional  spaces.  This  representation  can  generate  highly  detailed  outputs  while  having  parameters  with  relatively  small  dimensions.  Recently,  \cite{wolterink2022implicit} proposed  the  registration  method  based  on  the  implicit  neural  representation  of  displacement  field  estimation.  It  uses  Multi  Layer  Perceptron  (MLP)  to  establish  the  mapping  relationship  between  position  and  deformation,  optimizing  it  with  similarity  and  regulation.  \cite{han2023diffeomorphic}  introduced  a  Velocity  field  estimation-based  registration  method  of  implicit  neural  representation.But the velocity field in the \cite{han2023diffeomorphic} is not changing over time.

The  IGRT  is  an  essential  technology  for  precise  radiation  transfer.  In  the  domain  of  IGRT,  CBCT  is  frequently  used  due  to  its  rapid  acquisition\cite{letourneau2005cone, fu2020cone, ding2007characteristics},  cost  effectiveness  and  low  radiation  dose  advantages.  However,  CBCT  images  usually  exhibit  a  lower  contrast  for  soft  tissues,  limited  anatomical  details,  and  increased  artifacts  due  to  reduced  radiation  and  insufficient  projection  data.  In  contrast,  CT  images  provide  superior  voxel  value,  noise  reduction  and  spatial  uniformity,  making  them  more  suitable  for  treatment  planning.Therefore we need to align the planned CT to the CBCT to provide more detail and structure to the physician.

The  contributions  of  this  paper  are  as  follows:

(i)  Proposed  the  first  time-space  correlated  velocity  field  estimation  implicit  neural  representation  by  integrating  Implicit  Neural  Representation  with  LDDMM.

(ii)  Optimized  the  speed  of  INR-LDDMM  with  a  coarse-to-fine  framework.

(iii)  Validated  on  a  paired  CT-CBCT  interstitial  lung  disease  dataset  from  50  patient  prognoses,  achieving  state-of-the-art  results.

\section{Method}
In the proposed approach, we incorporate a Multilayer Perceptron (MLP) consisting of three layers into the Large Deformation Diffeomorphic Metric Mapping (LDDMM) framework. Initially, we formalize our problem (Section 2.1), followed by the presentation of the INR-LDDMM framework (Section 2.2). Finally, based on INR-LDDMM, we introduce improvements using a Coarse-to-Fine method (Section 2.3).

\subsection{Problem Formulation}
In the field of medical image registration, the Large Deformation Diffeomorphic Metric Mapping (LDDMM) is a fluid-based image registration model\cite{beg2005computing}, that calculates the spatial transformation $\varphi$ by computing a spatio-temporal velocity field $v(t, x)$. This is achieved through integration $\partial_t \varphi (t, x) = v(t, \varphi(t, x)),~ \varphi(0, x)=x$. For an appropriately regularized velocity field, it guarantees the diffeomorphic form transformation\cite{dupuis1998variational}. In medical imaging tasks, we need to align the moving image to a fixed image. We denote the moving image of size $a \times b \times c$ as $I_1$, and $I_2$ represents the fixed image of the same size. The underlying LDDMM optimization problem can be written as:

\begin{equation}
\mathcal L = \underset{v}{\text{argmin}}~   {Sim}(I_1(1),I_2)+\frac 12 \int_0^1 \|v(t) \|^2_L  dt ,
\end{equation}

satisfying $\partial_t I + \langle \nabla I, v\rangle=0,~ I_1(0)=I_1$, where $I_1(t)$ represents the moving image over time, with $I_1(1)$ being the final moved image. $\nabla$ signifies the gradient, $\langle \cdot, \cdot \rangle$ denotes the inner product and $Sim(A,B)$ signifies the similarity measurement between images. We assume the displacement field from $I_1$ to $I1(t)$ is $S$. Thereby, the optimization problem is mathematically presented as:

\begin{equation}
\underset{v}{\arg\min}\ \mathrm{Sim}(I_1 \circ S, I_2) +  \frac 12 \int_0^1 \|v(t) \|^2_L dt,
\end{equation}

\begin{figure*}[!h]
%\hspace*{-1cm}
\centering
\includegraphics[width=0.8\textwidth]{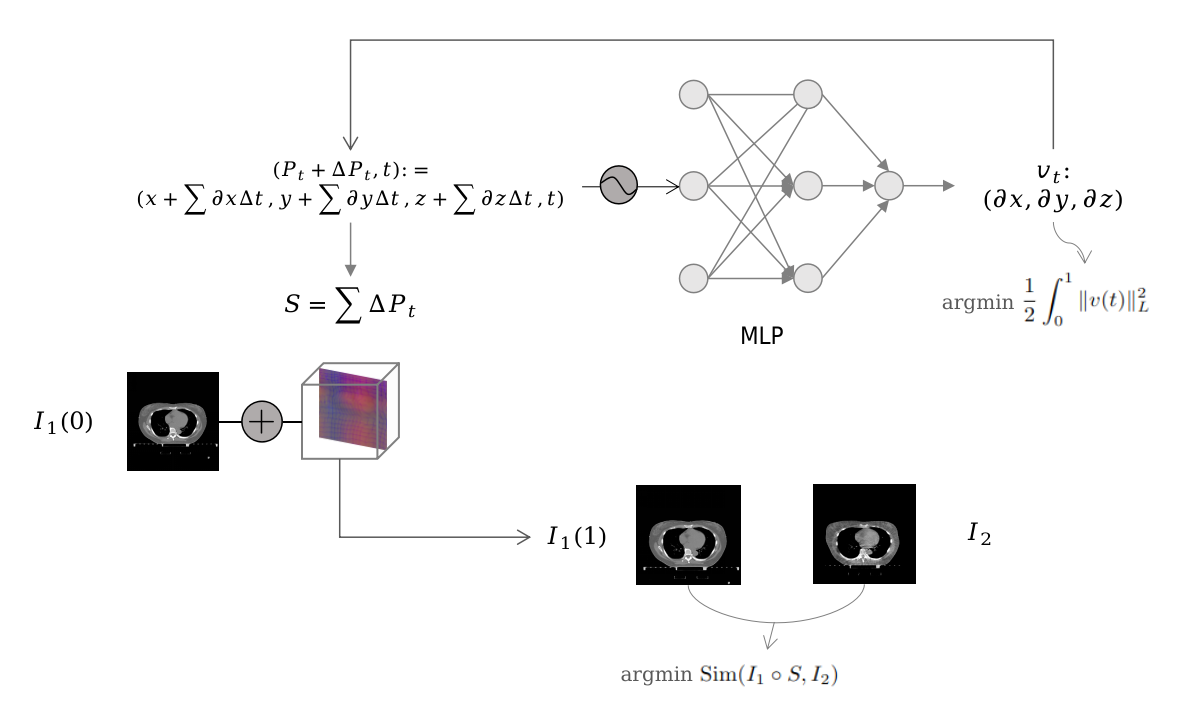}
\centering
\caption{Illustration of INR-LDDMM. We input the time t and the position coordinates corresponding to moment t into the MLP network to obtain the velocity at the corresponding moment. Based on the velocity we can get the amount of change in position and thus input the position and time of the next moment to predict the velocity of the next moment. According to the LDDMM model, our optimization objectives are the velocity and the similarity between the moved image and the fixed image.}
\label{fig:framework}
\end{figure*}

\begin{figure*}[!h]
%\hspace*{-1cm}
\centering
\includegraphics[width=0.8\textwidth]{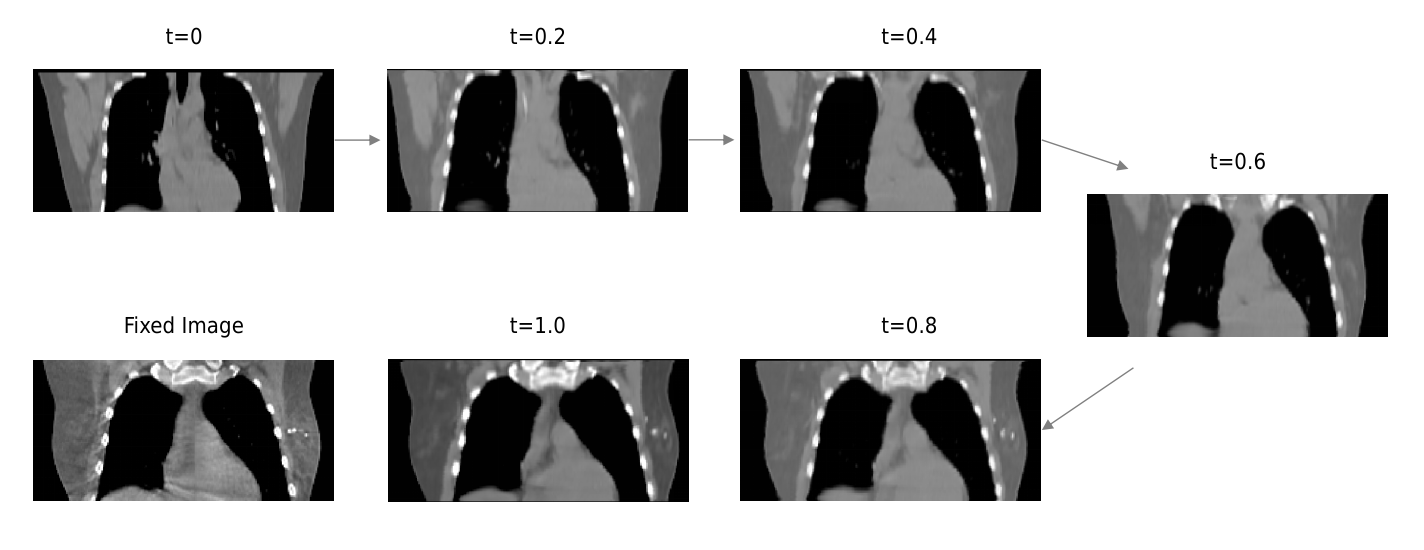}
\centering
\caption{The figure represents the transformation of the moving image over time. It can be seen that the moving image gradually and 'naturally' turns into a moved image over time.}
\label{fig:framework}
\end{figure*}

\begin{table*}[ht]
    \centering
    \begin{tabular}{l l l l l l l l l l}
    \hline
        Type & Method & CTV & Heart & Left Lung & Right Lung & Spinal & Aorta & Scapula & Average \\  \hline
        \multirow{2}*{Learning-based} & TransMorph & 0.83 & 0.82 & 0.90 & 0.89 & 0.67 & 0.76 & 0.73 & 0.80 \\ 
         & VoxelMorph & 0.79 & 0.78 & 0.89 & 0.90 & 0.62 & 0.73 & 0.73 &  0.77\\ 

        \multirow{2}*{Iteration-based} & Demos & 0.89 & 0.88 & 0.93 & 0.93 & 0.69 & 0.80 & 0.79 & 0.84\\ 
         & Elastix & 0.91 & 0.92 & 0.94 & 0.95 & 0.77 & 0.86 & 0.80 & 0.88\\ 

        \multirow{2}*{INR-based} & INR only & 0.90 & 0.92 & 0.95 & 0.95 & 0.77 & 0.87 & 0.81 & 0.87\\ 
        & Ours & \textbf{0.92} & \textbf{0.95} & \textbf{0.97} & \textbf{0.97} & \textbf{0.79} & \textbf{0.88} & \textbf{0.84} & \textbf{0.91} \\      
    \hline
    \end{tabular}
    \caption{Comparison of results with other methods. The methods are classified as Learning-based, Iteration-based and INR-based. The values in the table are DSC values.}
    \label{tab:comparison}
\end{table*}

where $\mathrm{Sim}(I_1 \circ S, I_2)$ represents the similarity measure between the transformed moving image and the fixed image.
\subsection{INR-LDDMM}
The INR-LDDMM framework is schematically shown in Fig.1.
We initially commence with a three-layered Multilayer Perceptron (MLP)  network. The design and development of this MLP network aims to replace  conventional velocity field estimation within the Large Deformation  Diffeomorphic Metric Mapping (LDDMM). Specifically, it assists us in  predicting local velocities at any given position within 3D images. We  define $\mathcal{X} = (x, y, z, t)$ as spatial positions and time  variables continuously inputted to MLP. The corresponding time velocity  is the output of the network, denoted as $\mathcal{V} = (v_x, v_y,  v_z)$. It is recognized that the relationship between displacement in  space $(\Delta x,\Delta y,\Delta z)$ and velocity $\mathcal{V} = (v_x,  v_y, v_z) = (\partial x,\partial y,\partial z)$ is represented as:
$$
 (\Delta x,\Delta y,\Delta z) = (\int v_x dt,\int v_y dt,\int v_z dt)
$$
 This equation has a discrete representation, thus the input $\mathcal{X} = (x+\Delta x, y + \Delta y, z + \Delta z, t)$ can be denoted as
 
\begin{align}
 (P_t + \Delta P_t,t) = (x+ \partial x\Delta t, y+\partial y\Delta t, z+\partial z\Delta t, t)
\end{align}

Consequently, for a given time interval $\Delta t$, after obtaining the  corresponding velocities $(\partial x,\partial y,\partial z)$ through  our velocity estimation MLP, we next determine the new input  $\mathcal{X} = (x+\partial x\Delta t, y+\partial y\Delta t, z+\partial z\Delta t, t)$. If we use $f(\mathcal{X})$  to represent MLP, this process is formulated as:
 
\begin{align}
 \mathcal{V} = f(\mathcal{X})
 \end{align}
 \begin{align}
 \mathcal{X} = (P_t + \Delta P_t, t)
  \end{align}
 \begin{align} 
 \Delta P_t = \mathcal{V} * \delta t = f(\mathcal{X}) * \delta t
\end{align}

 We can then obtain the displacement field:
 
$$
 S = \Sigma \Delta P_t= \Sigma [(\partial x\Delta t, \partial y\Delta t, \partial z\Delta t)]
$$

 Based on the preceding optimization goal, our current loss function to optimize is:

\begin{align}
L =& \underset{S}{\mathrm{argmin}} (Sim(I_1 \circ S, I_2) + \frac{1}{n}\sum_{i=1}^n \|v_i\|^2_L) \\
   =& \underset{S}{\mathrm{argmin}} (Sim(I_1 \circ \sum_{t=1}^n f(\mathcal{X}_t) * \delta t, I_2) \\
   &+ \frac{1}{n}\sum_{t=1}^n \|f(\mathcal{X}_t)*\delta t\|^2_L)
\end{align}

 Here, $\mathcal{X}_t$ represents the input at each step, encompassing  position and time, and $f(\mathcal{X}_t)$ indicates the output from the  MLP network at the corresponding time, that is, the velocity. The  $\circ$ and $\sum$ operators within the $Sim$ term designate the warping operation and the accumulation across all time-steps respectively.

We then utilize gradient descent to update the parameters of the MLP  network by solving the gradient of the loss function $L$. Iterating over this procedure continuously until satisfaction of a stopping criterion  (e.g., achieving a predetermined number of iterations or loss reduction  to a certain threshold), we finally return the resultant displacement  field $S$, which plays a vital role in transforming from the moving  image $I_1$ to the stationary image $I_2$.

\subsection{Coarse-to-fine Framework}
The direct optimization of high-resolution velocity fields could escalate the complexity of MLP optimization in proportion to the resolution. This tendency might trap the network in local optima more easily, besides prolonging the training duration. Consequently, we integrate a coarse-to-fine training mechanism into the INR-LDDMM framework.

We initiate the optimization at a coarse stage. The velocity field resolution at this stage, denoted as $(m_x, m_y, m_z)$, enables us to generate a rough displacement field, $S_{1}$, through the INR-LDDMM algorithm,
\begin{equation}
S_1 = \text{INR-LDDMM}((m_x,m_y,m_z),(I_1,I_2),n,MLP_{\text{param0}})
\end{equation}
where $MLP_{\text{param0}}$ represents the initially set MLP parameters, and $n$ denotes the iterations number. We define the resolution of the fine-stage velocity field as $(M_x,M_y,M_z)$. First, we need to interpolate $S_1$ from the low resolution $(m_x, m_y, m_z)$ to a high resolution $(M_x, M_y, M_z)$ via bilinear interpolation. We denote the interpolation results as $S_{1}’$ and utilize them for MLP network parameter optimization. The optimization objective is given by
\begin{equation}
L = {\mathrm{argmin}} \left\{ \|S - S_1'\|^2_L+ \frac{1}{n}\sum_{i=1}^n \|v_{i}\|^2_L\right\}
\end{equation}
where $S = \Sigma [(\partial x\Delta t, \partial y\Delta t, \partial z\Delta t)]$. Subsequently, we transfer the optimized MLP parameters, $MLP{\text{param}}$, into the fine-stage INR-LDDMM. It is given by:
\begin{equation}
S_f = \text{INR-LDDMM}(M_x,M_y,M_z,(I_1,I_2),n , MLP_{\text{param}})
\end{equation}
The final displacement field we aim for is $S_f$. The moved image $I_{1}(1)$ that we ultimately wish to obtain is:
\begin{equation}
I_{1}(1) = I_1(0) \circ S_f
\end{equation}

\section{Results}

\subsection{Dataset}
In this analysis, we studied a cohort of 50 patients who had received radiotherapy post-breast-conserving surgery. The study design involved paired CT-CBCT data. For each patient, we adhered to the standard treatment planning procedure, necessitating the procurement of CT images and at least one set of CBCT images throughout their treatment course.Initial cropping was done for CT and CBCT to ensure that they were in the same range. Both CT-CBCT images were 256*256*96 in size. Both CT and CBCT were labeled with segmentation masks, including CTV , Heart , Left Lung , Right Lung, Aorta, Scapula and Spinal Cord.

\subsection{Experiments}
We compare with deep learning based methods and traditional methods, and find that INR-LDDMM substantially leads other algorithms. The learning-based methods' results are averaged over five-fold cross-validation. The quantitative results are shown in Table 1. As shown in the figure, we show the results of the velocity field acting on the moving image for different time periods. It can be seen that the moving image is 'naturally' turned into a moved image at different time points.

\begin{algorithm}
\caption{INR-LDDMM for Medical Image Registration}
\label{INR-LDDMM}
\begin{algorithmic}[1]
\STATE Intialize three-layer MLP network 
\STATE Define $I_1$ as moving image and $I_2$ as fixed image.
\STATE Define $n$ as the time-step interval
\STATE Define $N_x$,$N_y$,$N_z$ as the density of the displacement field in the x, y, and z directions.
\FOR {each position $p = (x, y, z)$ in $Meshgrid(0:1:N_x,0:1:N_y,0:1:N_z)$, and time $t$}
\STATE Predict velocity $v_p = (\delta x, \delta y, \delta z)$ using MLP network
\STATE Update position $p$ as $p + v_p \delta t$
\STATE Update time $t$ as $t - \delta t$
\IF {time step divisible by $n$}
\STATE Save velocity $v_p$ as $v_{i|p}$, where $i = \lfloor t / n \rfloor$
\ENDIF
\ENDFOR
\STATE Save velocity $v_{i|p}$ in every $p$  as velocity field $v_{i}$.
\STATE Compute displacement field $S$ as average of $v_1 t, v_2 t, \ldots, v_n t$
\STATE Compute loss as:
\[
L = \underset{S}{\mathrm{argmin}} \left\{ Sim(I_1 \circ S, I_2) + \frac{1}{n}\sum_{i=1}^n \|v_i\|^2_L\right\}
\]

\STATE Optimize MLP network using gradient descent or another optimizer to minimize $L$
\IF {stopping criterion fulfilled}
\STATE Exit loop
\ELSE
\STATE Return to step 5
\ENDIF
\RETURN displacement field $S$
\end{algorithmic}
\end{algorithm}

\begin{algorithm}
\caption{Coarse-to-Fine INR-LDDMM for Medical Image Registration}
\label{coarse-to-fine}
\begin{algorithmic}[1]
\STATE Initialize three-layer MLP network
\STATE Set moving image $I_1$, fixed image $I_2$, time-step interval $n$, density of fine displacement field $(M_x,M_y,M_z)$ and density of coarse displacement field $(m_x,m_y,m_z)$.

\STATE Define INR-LDDMM(($N_x$,$N_y$,$N_z$),($I_1$,$I_2$),n,$MLP_{param}$) as Algorithm 1.($N_x$,$N_y$,$N_z$) is density of the displacement field, ($I_1$,$I_2$) is image pair of fixed image and moving image,n is the time-step interval and $MLP_{param}$ is incoming parameters of MLP.
\STATE $S_1$ = INR-LDDMM(($m_x$,$m_y$,$m_z$),($I_1$,$I_2$),n,$MLP_{param0}$)
\STATE Interpolate $S_1$ to the density of $(M_x,M_y,M_z)$, denoted as $S_1'$.
        
\FOR {each $p = (x, y, z)$ and $t$}
\STATE Predict velocity $v_{i|p} = (\delta x, \delta y, \delta z)$ using MLP network
\STATE $p \leftarrow p + v_{i|p} \delta t$
\STATE $t \leftarrow t - \delta t$
\IF {time step divisible by $n$}
\STATE Save velocity $v_{i|p}$, where $i = \lfloor t / n \rfloor$
\ENDIF
\ENDFOR
\STATE Save velocity $v_{i|p}$ in every $p$ as velocity field $v_{i}$.
\STATE Compute displacement field $S$ as average of $v_{1} t, v_{2} t, \ldots, v_{n} t$
\STATE Compute loss as:
\[
L = {\mathrm{argmin}} \left\{ \|S - S_1'\|^2_L+ \frac{1}{n}\sum_{i=1}^n \|v_{i}\|^2_L\right\}
\]
\STATE Optimize MLP network using gradient descent or another optimizer to minimize $L$
\IF {stopping criterion fulfilled}
\STATE Exit loop
\ELSE
\STATE Return to step 6
\ENDIF
\RETURN  $MLP_{Param}$.
\STATE $S_f$ = INR-LDDMM($M_x$,$M_y$,$M_z$,($I_1$,$I_2$),n , $MLP_{param}$).
\RETURN displacement field $S_f$
\end{algorithmic}
\end{algorithm}

\section{Conclusion}
In this study, we proposed a novel registration algorithm named INR-LDDMM that integrates the effectiveness of Implicit Neural Representation into the Large Deformable Diffeomorphic Metric Mapping method.With extensive experiments on datasets consisting of paired CT-CBCT images, INR-LDDMM consistently outperformed traditional and deep learning-based medical image registration methods.

\section{Compliance with Ethical Standards}
This retrospective study was approved by the ethics committee of Ethics Committee of Sun Yat-sen University Cancer Center (SL-G2023-052-01).

\section{Acknowledgments}
This work is partly supported by grants from the National Natural Science Foundation of China (82202954, U20A201795, U21A20480, 12126608) and the Chinese Academy of Sciences Special Research Assistant Grant Program.
% References should be produced using the bibtex program from suitable
% BiBTeX files (here: strings, refs, manuals). The IEEEbib.bst bibliography
% style file from IEEE produces unsorted bibliography list.
% ------------------------------------------------------------------------- 
\bibliographystyle{IEEEbib}
\bibliography{strings,refs}

\end{document}